\documentclass{PoS}

\usepackage{aas_macros}

\usepackage{lineno}

\title{Latest news from the HAWC outrigger array }

\ShortTitle{Latest news from the HAWC outrigger array}

\author{\speaker{Vincent Marandon} \\ 
        Max-Planck-Institut f\"ur Kernphysik\\
        E-mail: \email{vincent.marandon@mpi-hd.mpg.de}}

\author{ Armelle Jardin-Blicq \\ 
        Max-Planck-Institut f\"ur Kernphysik\\
        E-mail: \email{armelle.jardin-blicq@mpi-hd.mpg.de}
}
\author{ Harm Schoorlemmer \\ 
        Max-Planck-Institut f\"ur Kernphysik\\
        E-mail: \email{harmscho@mpi-hd.mpg.de}
}

\author{for the HAWC Collaboration\footnote{for collaboration list and acknowledgements see PoS(ICRC2019)1177 and https://www.hawc-observatory.org/collaboration/icrc2019.php}\\
     }

\abstract{The High Altitude Water Cherenkov (HAWC) observatory is a very high energy gamma-ray detector located in Mexico. In late 2018, the HAWC collaboration completed a major upgrade consisting of the addition of a sparse outrigger array of 345 small water Cherenkov detectors (WCDs) surrounding the 300 WCDs of the main array and extending the instrumented area by a factor of 4. It provides an improved reconstruction of the showers whose core and footprint are not well contained in the array and increases the effective area in the range of a few TeV to beyond 100 TeV. This improvement in sensitivity will help to have a better understanding of the Galactic sources that accelerate particles up to the knee of the cosmic ray spectrum. In this contribution, we will show the current status, the performance, and the first results from the HAWC outrigger array. }

\FullConference{36th International Cosmic Ray Conference -ICRC2019-\\
		July 24th - August 1st, 2019\\
		Madison, WI, U.S.A.}

\begin{document}

\section{Introduction}

The High Altitude Water Cherenkov (HAWC) is a very-high-energy gamma-ray observatory located at 4100~m altitude on the volcano Sierra Negra in the state of Puebla in central Mexico.
It is composed of 300 water Cherenkov detectors (WCD). Each of them is a cylinder of 7.3~m in diameter and 4.5~m in height. The array covers a total area of $\mathrm{\sim22000~m^2}$.
They are equipped with 4 upward facing photomultiplier tubes (PMT), one 10'' Hamamatsu R7081-HQE at the centre of the tank and three 8'' R5912 halfway to the edges.
The observatory is sensitive to gamma rays ranging from a few hundreds GeV up to $\sim$100 TeV, with an instantaneous field of view of $\mathrm{\sim2sr}$ and a duty cycle of more than 95\%.
Those two characteristics compensate the smaller instantaneous sensitivity compared to the imaging atmospheric Cherenkov technique, especially at the highest energies for which fluxes tend to be low.
This makes this instrument very well suited to the hunt for sources that can accelerate particles up to the knee of the cosmic ray spectrum ($\mathrm{\sim10^{15}~eV}$).

Around 10 TeV, the footprint on the ground produced by a gamma-ray shower is comparable to the size of the instrument.
Therefore, most of the events triggering the instrument will fall outside the main array which makes the event reconstruction less accurate because of degeneracy between the core position and the energy of the shower.
In order to improve the reconstruction of those events thus improving the instrument sensitivity, a sparse outrigger array of small tanks has been added around the main array.

\section{Array Layout}

\begin{figure}
\begin{center}
\includegraphics[width=0.41\textwidth]{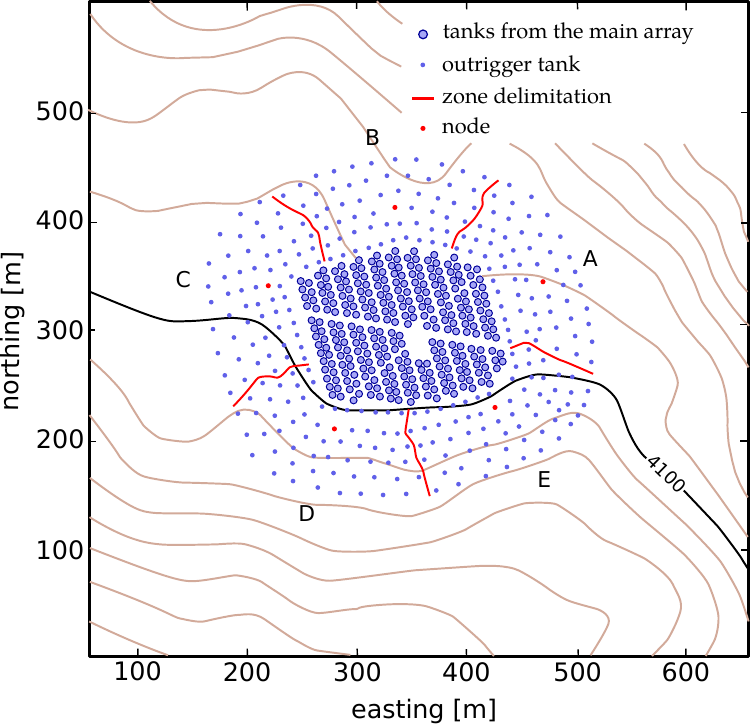}
\includegraphics[width=0.58\textwidth]{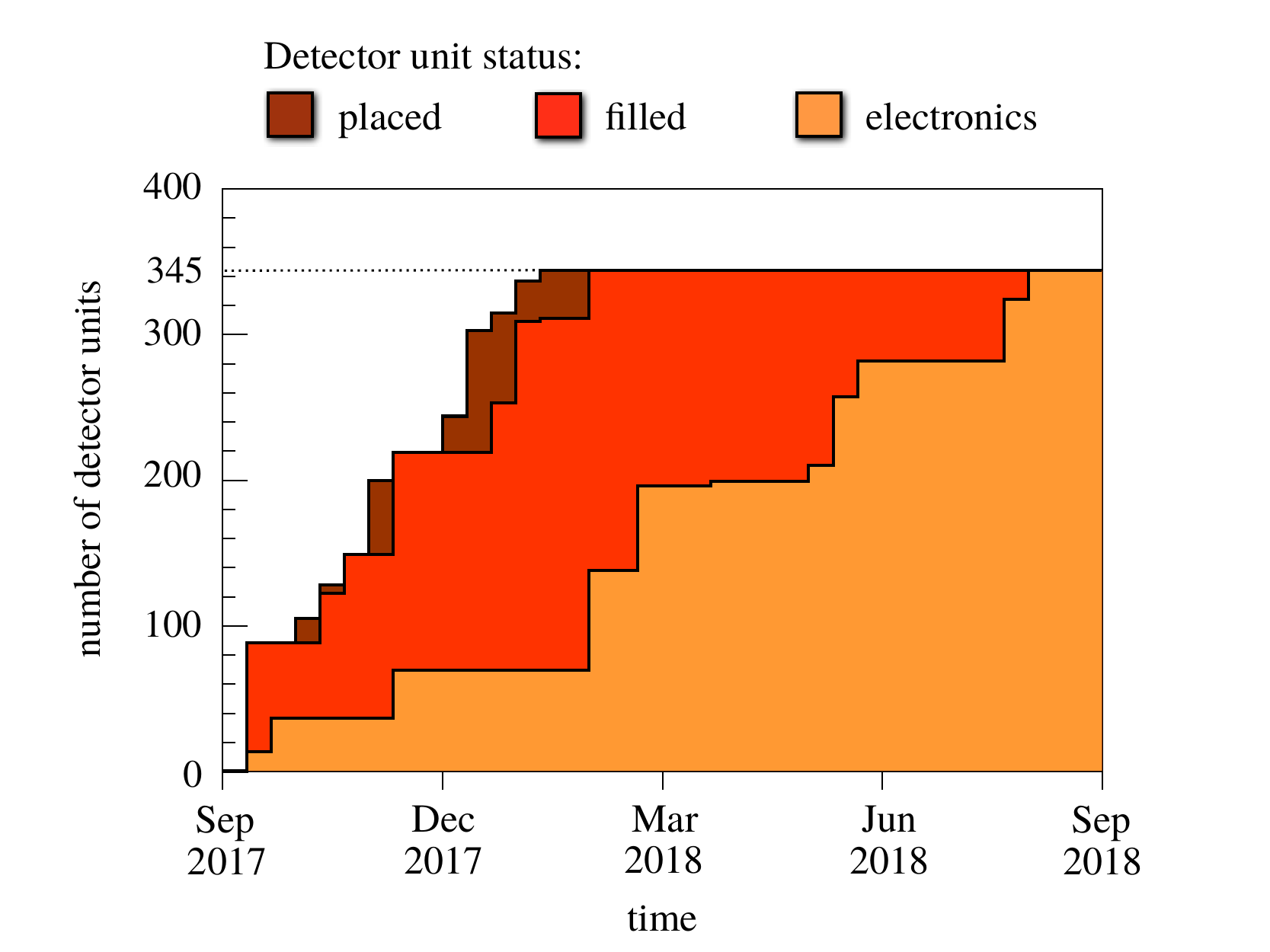}
\caption{ (\textit{left}) Top view sketch of the HAWC main and outrigger array. The 5 sections are delimited by red lines and named by letters from A to E.
The red dots represent the nodes that contain the readout electronics and power supply for each tanks.
(\textit{right}) Outrigger deployment timeline.}
\label{fig::layout}
\end{center}
\end{figure}

The outrigger array is composed of 345 tanks filled with purified water.
Each tank is 1.55~m in diameter and 1.65~m in height and they are separated with distance between 12 to 18~m. 
They are equipped by a single Hamamatsu R5912 8'' photomulitplier tube (PMT) upward facing and anchored at the bottom of the tank.
In order to limit the cable length between tanks and the readout electronics, the array is divided into 5 sections, named with letters from A to E (see Fig~\ref{fig::layout}) containing 69 tanks each.
The equipment needed to run and acquire the data for each section is centralised into nodes located in the middle of each section.
In order to keep the time synchronised between the different nodes and the main array, a White Rabbit system \cite{WhiteRabbit} is used which allows a sub-nanosecond level precision.

The deployment of the outrigger array happened between September 2017 and September 2018 (the timeline can be seen in Figure~\ref{fig::layout}). 
The array has been continuously taking data since then.

\section{Readout Electronics}
Analogue signals from the PMTs are sent through RG-59 cables to the node.
This signal is separated from the HV that powers the PMT by a pick-off module and sent to the readout electronics via Ethernet cables.
For each section, the readout electronics consists of three Flash Analogue-to-Digital Converter (FADC) boards.
They convert the analogue signal to a 12 bit digital signal with a sampling rate of 250 MHz rate for 24 channels.
These boards have originally been developed for the FlashCam project which is a prototype of camera proposed for the Cherenkov Telescope Array (CTA) \cite{2015ICRC...34.1039P}.
Among those 24 channels, 23 are used for tank signals and the last channel is used to trigger the readout during calibration with a laser.

\section{Trigger conditions and data merging with the main array}

\begin{figure}
\begin{center}
\includegraphics[width=0.8\textwidth]{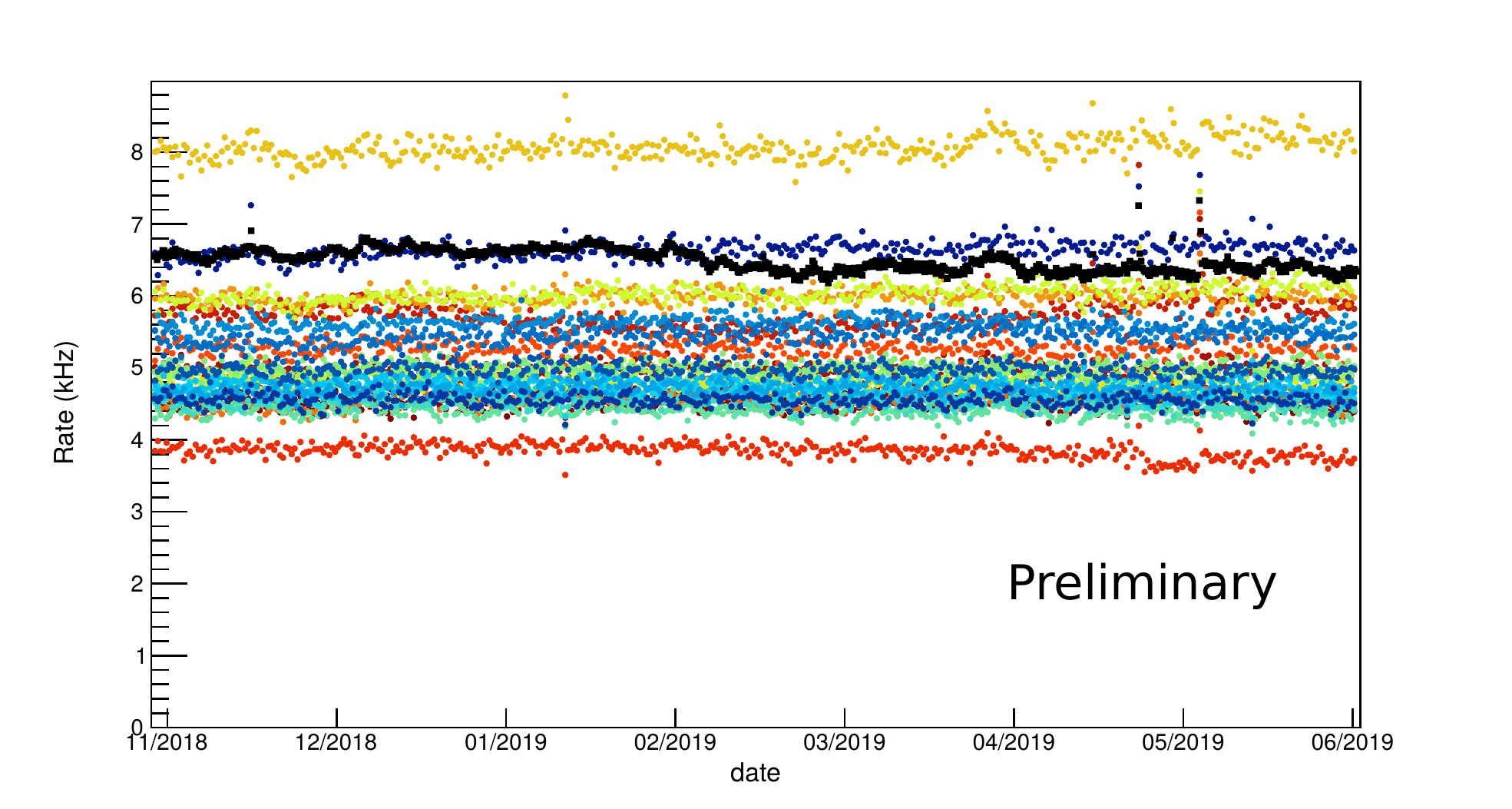}
\caption{The coloured markers indicate the measured individual trigger rate for the 23 channels of one FADC over 7 months. Each point is an average over 12 hours. The black squares represent the acquisition rate.}
\label{fig::rateevo}
\end{center}
\end{figure}

The trigger decision is taken by the FPGA on board each FADC card, independently from the others.
This implies that the whole array is effectively divided into 15 independent sub-arrays.
In order to declare a trigger, at least 2 channels out of the 23 tank channels with a signal above 1 photo-electron (pe) within a time window of 160 ns are required.
When those conditions are met, the 24 channels are read out for 200 ns before and after the trigger time and the traces are sent via 10 GBit optical fibres to the main server where the data reduction is performed.
This step consists of a 4 stages multi-threaded pipeline that reads the information received from the board, discards the data from the empty channels, performs the pulse reconstruction for the remaining channels by using the calibration coefficients and stores this reduced information to disk.
In a second step, only the outrigger events that are in coincidence with the main array events are kept. The coincidence window in this case is 0.5~$\mathrm{\mu}$s before and 1~$\mathrm{\mu}$s after the main array trigger time.

Each outrigger PMT on site, as well as the system rate for each FADC are continuously monitored to check for the stability of the system and identify possible problems, like light leaks for example.
Figure~\ref{fig::rateevo} shows the evolution of the single PMT and acquisition rate, averaged over 12h, for 7 months of operation since the full deployment. 
Some spikes can be seen in the acquisition rate, which are related to days at which thunderstorms affects the system.
Otherwise, the system is very stable over time with fluctuations of the order of few percent on average.

\section{Calibration of the outrigger data} 

\begin{figure}
\begin{center}
\includegraphics[width=0.8\textwidth]{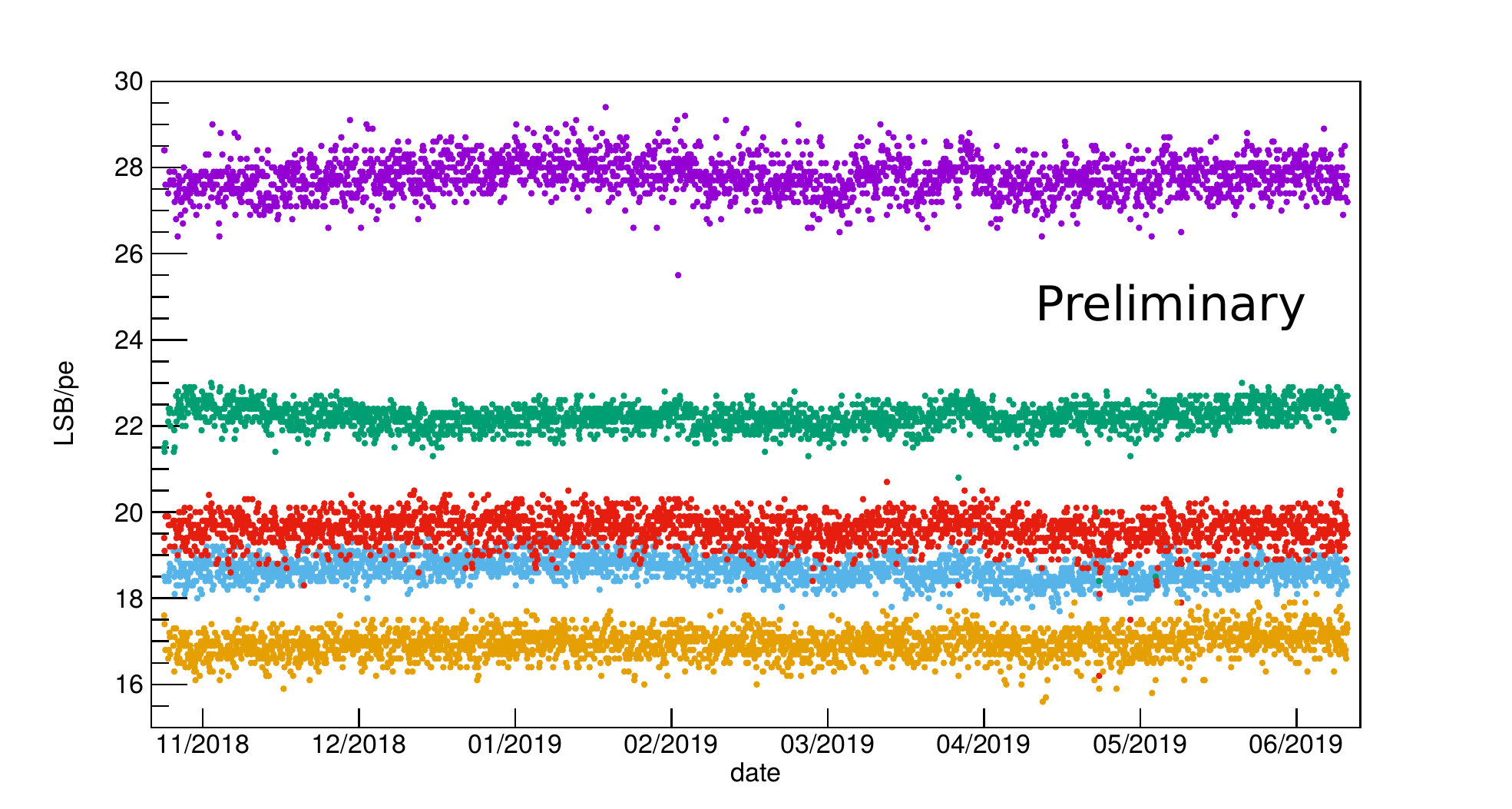}
\caption{Measured gain evolution over 7 months for 5 typical PMTs. The level of fluctuation is of the order of $\sim$1.5\%. The gain is evaluated roughly every 2 hours.}
\label{fig::gainevo}
\end{center}
\end{figure}

The calibration is performed via two methods using a laser and the air shower data.
Each tank is equipped with an optical fibre connected to a 532~nm central laser, which directly illuminates the PMT from above.
The laser system is equipped with filter wheels in order to get illumination from single pe to thousands of pe. 
The intensity delivered after attenuation is measured by a radiometer.
This information in combination with the fraction of pulses detected is used to derived the number of pe delivered to each PMT by the so-called \textit{occupancy method} \cite{HAWCCalibration}.

\begin{figure}[t]
\begin{center}
\includegraphics[width=0.6\textwidth]{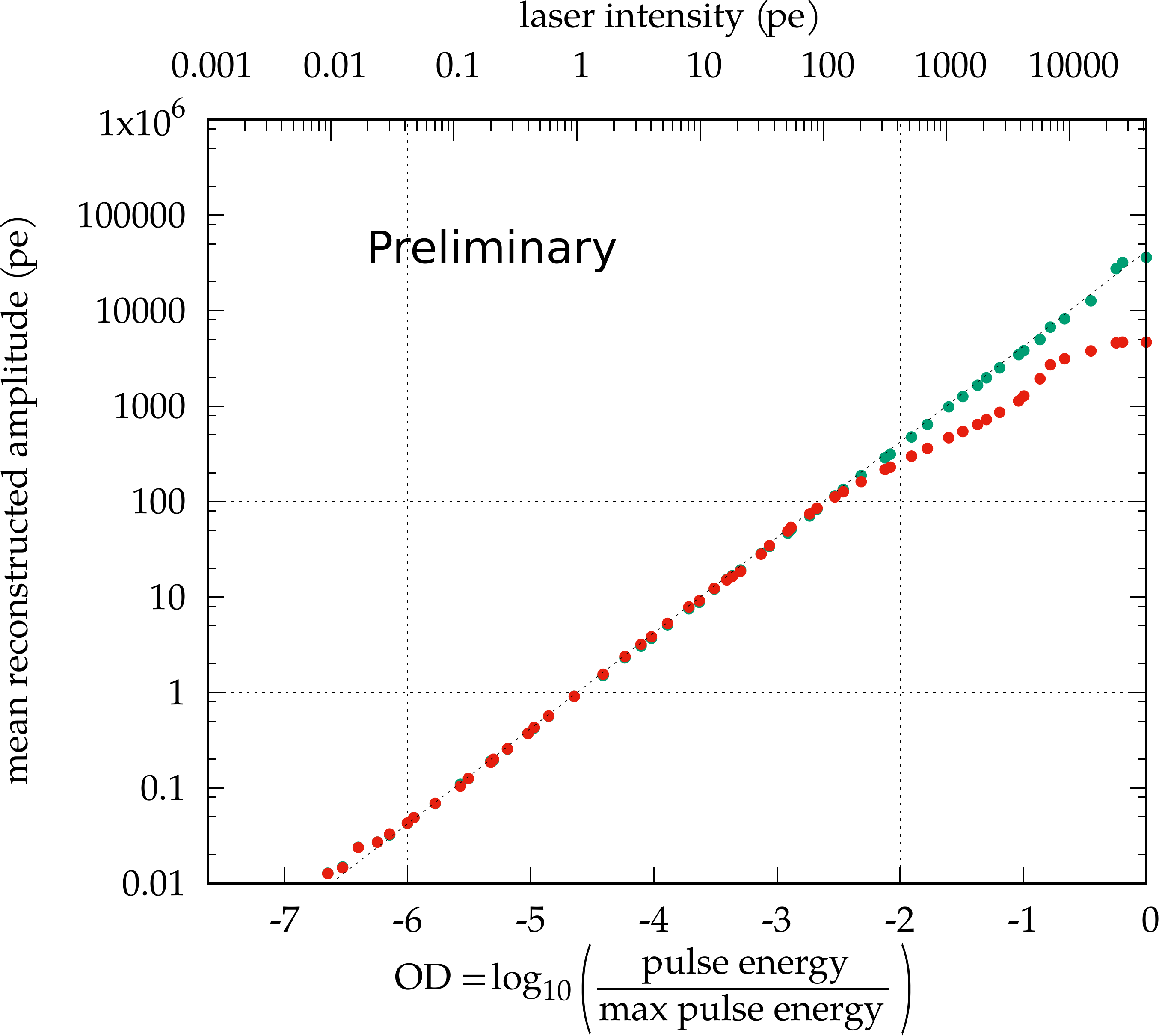}
\caption{This graph shows the reconstructed charge as a function of the laser intensity and the optical depth measured by a radiometer. 
The red points are derived by applying the current pulse reconstruction scheme. The green points are derived after applying the correction to linearity. The tilted dotted line shows the expectation for an ideal charge reconstruction.}
\label{fig::laserreco}
\end{center}
\end{figure}

\begin{figure}
\begin{center}
\includegraphics[width=1\textwidth]{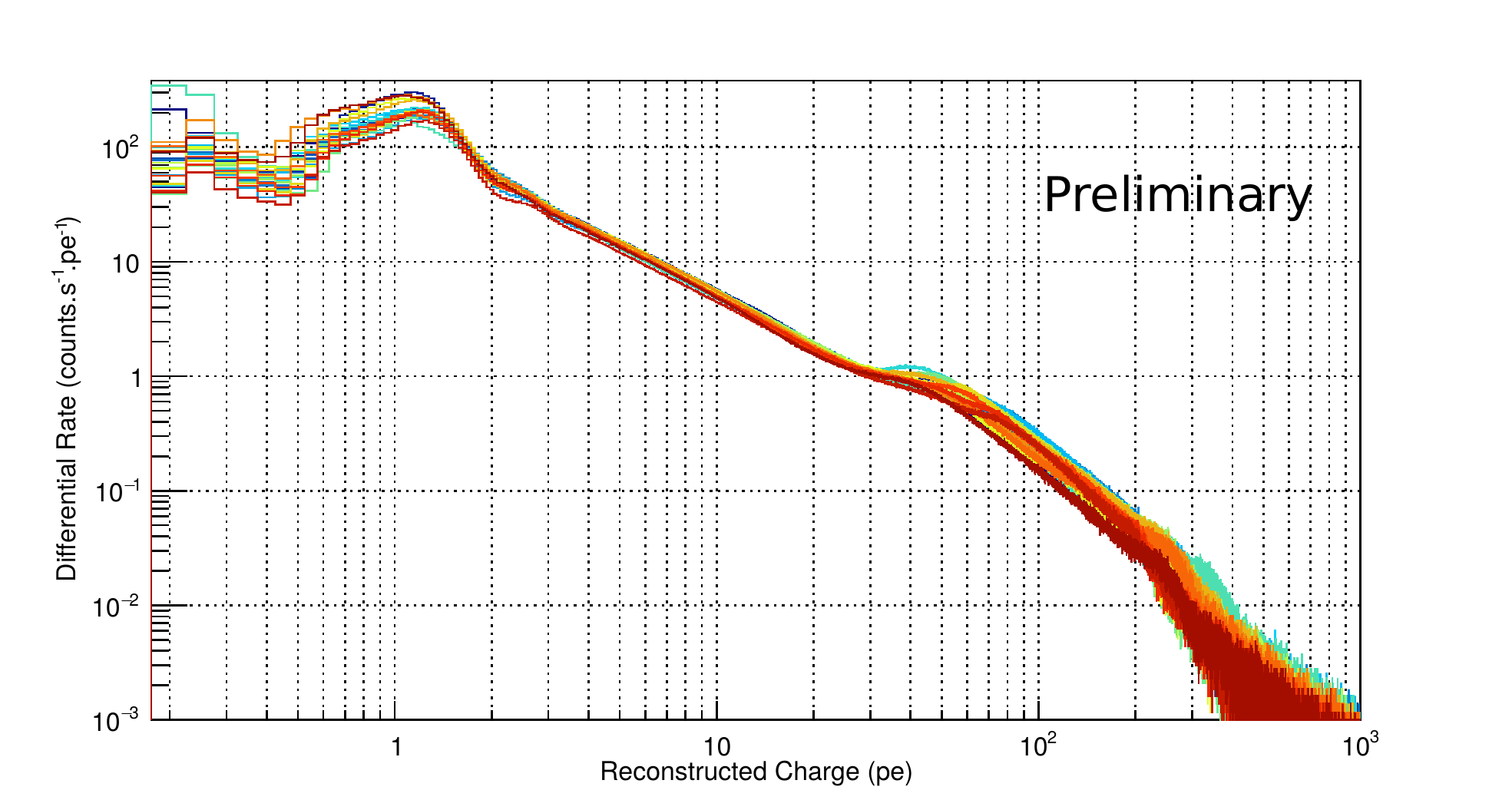}
\caption{Distribution of reconstructed charge for the 23 PMT of one FADC board. The linearity correction has not been applied.}
\label{fig::chargerate}
\end{center}
\end{figure}

The gain is derived using the signals acquired during regular data taking.
For each event, part of the waveform that is far from the trigger position is analysed.
Events that arrive in this late time are mainly dark counts or other events that are unrelated to the shower responsible for the trigger.
Pulses that are way above the single pe value are filtered out and a distribution is made with the remaining events which are used to derive the gain. 
This method requires few hours of acquisition in order to get accurate enough calibration coefficients.
An important advantage of performing the calibration on the data is to be able to actively monitor the evolution of the system. It was stable over the last 7 months, as can be seen on the Figure~\ref{fig::gainevo}.

The laser calibration system is used to check the charge reconstruction over a wide range of illumination and to derive timing corrections to apply to the data.
The comparison between the event charge reconstruction and the laser can be seen in Figure~\ref{fig::laserreco}. There is a good agreement between them, nevertheless, the deviation to the linear regime is getting important above $\sim$200 pe and requires a correction that is derived from the laser signal.
The reconstructed charge distribution for 23 channels of the same FADC board is shown in Figure~\ref{fig::chargerate}. This has been done on triggered events during regular data taking without the linearity correction. 
It shows a dispersion in the charge reconstruction of the order of 15-20\%, which could be due to differences in quantum efficiency between PMTs but also water level and quality.
The overall performance of the reconstruction is good. The charge reconstruction follows the Poisson limit up to the point for which the deviation to linearity becomes important. The time resolution obtained is $\sim$250 pico-seconds at 100 pe.

\begin{figure}
\begin{center}
\includegraphics[width=0.514\textwidth]{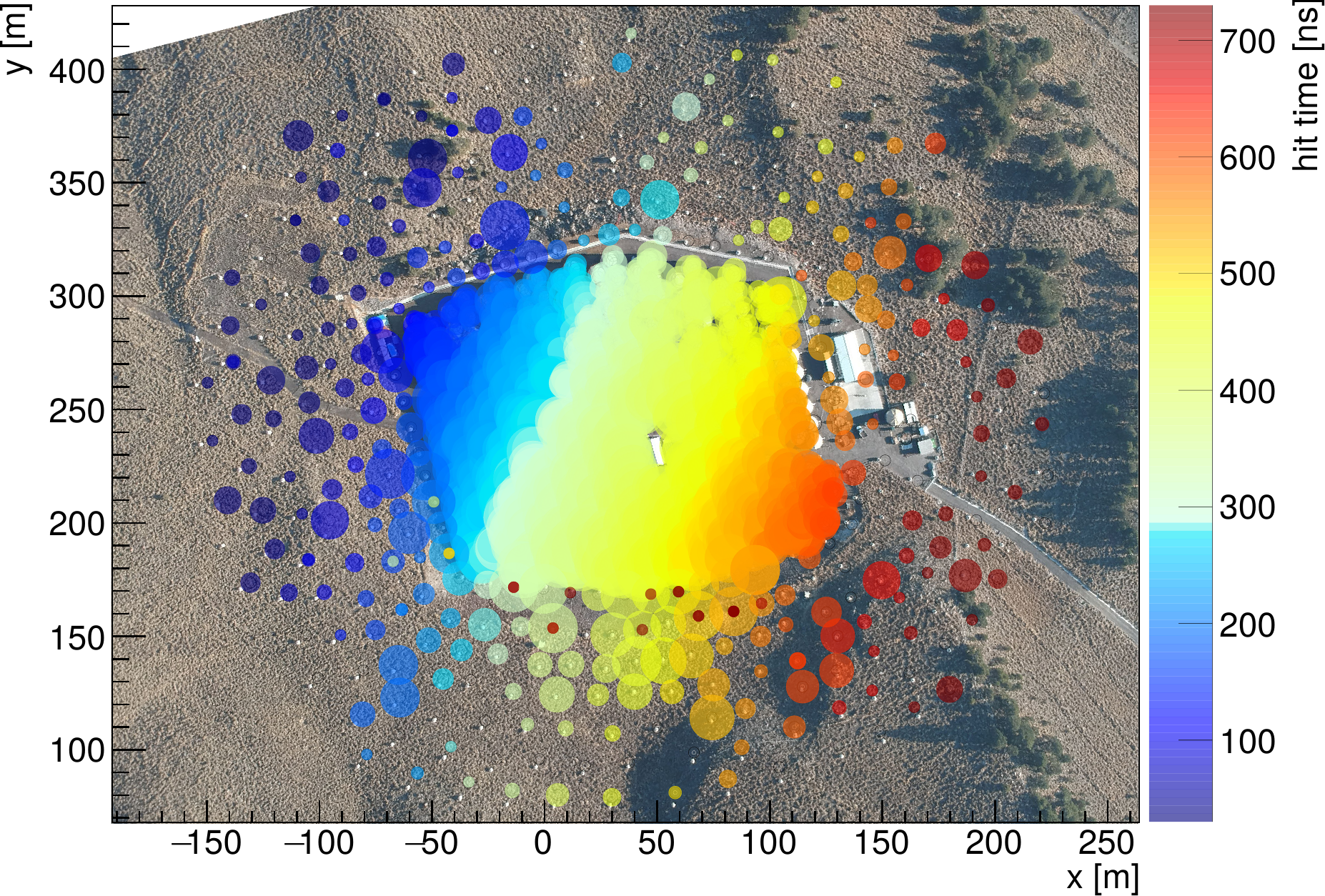}
\includegraphics[width=0.48\textwidth]{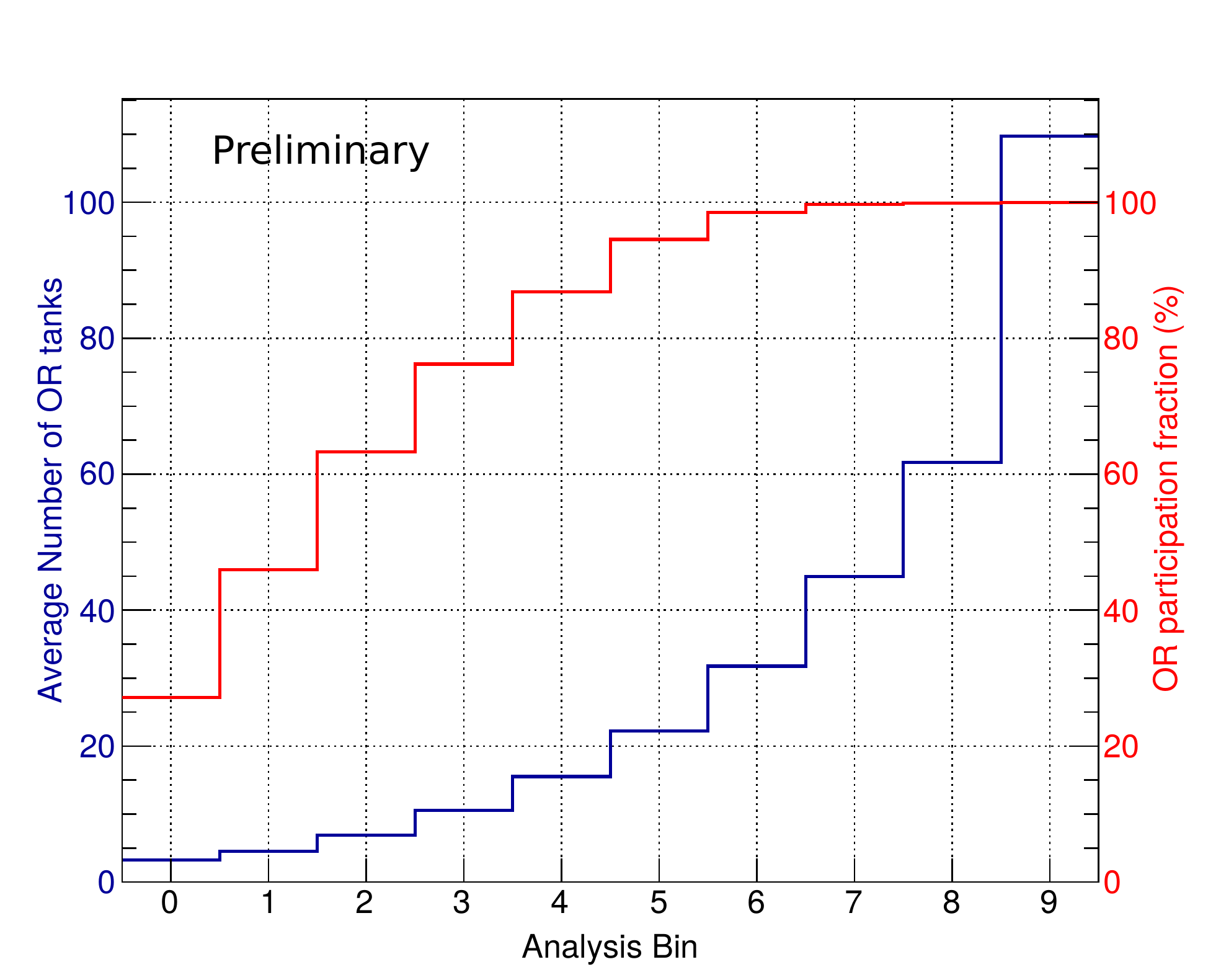}
\caption{(\textit{left}) Example of an air shower that triggered the main array and the outrigger array. The color scale represents the relative arrival time of the particles in the tanks. (\textit{right}) 
The red curve shows the outrigger array participation fraction to the main array events as a function of the Analysis Bin. The latter is the fraction of the main array tanks that participated in a shower event and is related to its energy (see \cite{CrabHAWC} for details).
The blue curve shows the average multiplicity of the outrigger array when it participates in a main array event.}
\label{fig::plotpart}
\end{center}
\end{figure}

\clearpage
\section{Conclusions and Outlook}

The outrigger array has been successfully deployed since September 2018 and is fully operational.
The array extension is running smoothly and the system is understood and stable.
The additional information provided by the outriggers is illustrated in Figure~\ref{fig::plotpart}; 
it shows the outrigger array participation fraction to events detected by the main array and the average multiplicity of the outrigger tanks when this array participates.
One can see that at the highest analysis bins, which correspond to the highest energies, the outrigger array always participates with, on average, a large number of tanks.
The next step for the project is to take this additional information into account in the event reconstruction.
As shown in \cite{ORCoreReco}, the expected improvement in core resolution for events that fall on the outrigger array is of the order of a factor 2-3.
This should improve the reconstruction of the other shower parameters, like energy and the angular resolution and finally improve the sensitivity of the instrument.

\end{document}